\documentclass[preprint,12pt,authoryear]{elsarticle}
\usepackage{amsmath}
\usepackage{graphicx}
\usepackage[svgnames]{xcolor}
\usepackage{natbib}
\usepackage[a4paper]{geometry}
\usepackage[textwidth=8em,textsize=small]{todonotes}
\usepackage{amsmath}
\usepackage{natbib}
\usepackage[utf8]{inputenc}
\usepackage{graphicx}
\usepackage{longtable}
\graphicspath{ {./images/} }
\usepackage{multirow}
\usepackage{float}
 \usepackage{xspace}
 \usepackage{appendix}
 \usepackage{array}
\usepackage{amssymb}
\usepackage{dirtytalk}
\usepackage{etoolbox}
\usepackage{url}
\newcolumntype{C}[1]{>{\centering\arraybackslash}p{#1}}

\usepackage{booktabs}
\usepackage{xspace,color}
\usepackage{url}
\usepackage[T1]{fontenc}
\usepackage[american]{babel}

\usepackage{graphicx}
\usepackage{fancyvrb}
\usepackage{listings}
\usepackage{bm}
\usepackage{xcolor}

\xdefinecolor{gray}{rgb}{0.4,0.4,0.4}
\xdefinecolor{blue}{RGB}{58,95,205}

\journal{Insurance: Mathematics and Economics}

\begin{document}

\begin{frontmatter}

\title{Mortality Forecasting with Generalized Additive Mixed Models}
\author{Reza Dastranj\corref{cor1}}

\ead{dastranj@math.muni.cz}

\author{Martin Kol\'a\v r}
\ead{mkolar@math.muni.cz}

\address{Department of Mathematics and Statistics, Masaryk University, Kotlářská 2, 611 37 Brno, Czech Republic}

\cortext[cor1]{Corresponding Author}

\begin{abstract}
This study introduces a novel generalized additive mixed model (GAMM) for mortality modelling, utilizing the mortality covariate $k_t$ as proposed by Dastranj-Kolar. Our findings indicate that the GAMM effectively addresses this shortcoming. Given that ASDRs constitute longitudinal data, as noted in the LME framework, the GAMM offers a more flexible and suitable approach for both modeling and forecasting mortality rates. Empirical evaluations using data from the Human Mortality Database (HMD) demonstrate the GAMM's strong ability to reproduce observed mortality patterns with high precision. Comparative analyses show that the GAMM consistently outperforms the LL model in both in-sample fit and out-of-sample forecasting across multiple populations. These results highlight the GAMM's potential as a robust and reliable tool for mortality modeling and long-term demographic forecasting.

\end{abstract}


\end{frontmatter}

\section{Introduction}\label{sec:intro}

The concept of human lifespan risks can be neatly categorized into two distinct types. The first type pertains to unsystematic mortality risk, which revolves around the uncertainty surrounding the actual number of deaths within a specific population compared to the expected count. This risk can be somewhat mitigated by expanding the population size under the assumption of disconnected individual lifetimes. However, unforeseeable events like widespread epidemics can lead to sudden and significant spikes in mortality that cannot be counteracted through diversification. The second category is systematic mortality risk or longevity risk, encompassing the uncertainty of future mortality rate deviations from predictions and expectations.

Longevity risk, the prospect of people living longer than anticipated, has far-reaching implications for individuals, pension schemes, and insurance companies. Individuals might outlive their savings, resorting to government assistance or charity. Pension schemes could find themselves obligated to pay more in benefits than planned, leading to financial hardships. Insurance companies might face unexpected payouts, causing financial losses. 
Longevity risk has surged to become a central concern for pension schemes, particularly when evaluating risk profiles amidst declining discount rates and rising liabilities.

Advances in medicine and lifestyle changes have contributed to increased lifespans. For instance, the United States' average life expectancy has grown by over 30 years since 1900. Some individuals possess genetic predispositions for longer lifespans.

 If there exists a maximum conceivable human lifespan, we have not yet come close to it (\citealt{mccarthy2023mortality}). By analyzing the lifespans of individuals in the UK, spanning birth years as far back as $1880$, \cite{mccarthy2023mortality} aim to extrapolate potential future scenarios. The projections indicate that men born in $1970$ might possibly live up to 141 years, while the eldest women from the $1970$ cohort could reach 131 years. Furthermore, our estimations demonstrate a $95\%$ likelihood that the final survivor among Swedish women born in 1950 may pass away between the ages of $117.1$ and $125.5$ years.

The imperative of developing predictive models to anticipate mortality rates takes on paramount importance in the effective management of longevity risk. Employing increasingly sophisticated mortality models significantly bolsters longevity risk management by providing more precise forecasts of future mortality rates. These predictive insights, in turn, empower stakeholders to exercise discerning and informed judgment in financial matters.

The adept application of advanced mortality models affords us meticulously calibrated projections of upcoming mortality rates. This predictive data serves as a valuable asset, guiding the formulation of prudent financial decisions. These decisions span critical determinations, including optimizing retirement savings strategies and assessing the feasibility of acquiring annuities. The strategic integration of advanced mortality models underscores their central role in refining methodologies aimed at addressing longevity risk. This strategic refinement fosters heightened financial security and facilitates a culture of well-considered, astute decision-making.

Recent mortality trends have shown substantial volatility, making accurate prediction of future longevity increasingly challenging. The uncertainty around lifespans urges trustees and sponsors to explore innovative strategies for managing longevity and mortality risk in various-sized pension schemes.

Accurate mortality projection models play a pivotal role in aiding individuals and financial institutions to comprehend the potential implications of extended lifespans on retirement plans. These models quantify the probability of surpassing expected ages, facilitating informed decisions regarding savings, investments, and retirement strategies.

Retirement savings are pivotal for individuals during their retired years. Mortality projection models provide insights into the necessary longevity of retirement funds, assisting in determining the required savings and investments for an extended retirement period.

Insurance companies, providers of products like annuities, rely on precise mortality models to set appropriate payout rates. The development of accurate mortality models empowers insurers to determine annuity payout rates, factoring in the risk of policyholders living longer than anticipated.

LME models are particularly valuable for analyzing longitudinal data, such as ASDRs collected over time from different populations. By combining both fixed effects and random effects, LME models enable the capturing of both population-level trends and individual-level variations within the data. Fixed effects help uncover general patterns and associations between mortality rates and various factors of interest, such as age and gender. On the other hand, random effects address the correlations and dependencies present within specific groups, accounting for the similarities among observations within the same age group. In the context of mortality forecasting, LME models offer several advantages and insights. Their ability to integrate fixed and random effects enables a comprehensive understanding of the underlying factors influencing mortality rates. They can capture both overall population-level trends and specific variations within subgroups. Additionally, LME models are well-suited for handling the longitudinal nature of mortality data collected sequentially over time. This allows for the incorporation of temporal dependencies and the exploration of trends and changes in mortality rates across different time periods.

\cite{Dastranj27042025} introduces an innovative approach using linear mixed-effects models for the analysis and prediction of mortality rates. This study's key contributions include the integration of age, the interaction between gender and age, and their interactions with the predictors $k_t$ and $k_{ct}$, and cohort as fixed effects. Furthermore, we have incorporated additional random effects to account for variations in the intercept, predictor coefficients, and cohort effects among different age groups of females and males across various countries. We conducted a comparative analysis by assessing the performance of the LME model against the LC models fitted to individual populations. Additionally, we evaluated the predictive accuracy of the LME model in comparison to the LL model. Our findings demonstrate that the LME model offers a more precise representation of observed mortality rates within the Human Mortality Database (HMD). It exhibits robustness in selecting calibration rates and outperforms the LC and LL models. These results highlight the suitability of the LME model as a framework for modeling and forecasting mortality trends. Therefore,  \cite{Dastranj27042025}  highlights the significance of utilizing LME models in mortality forecasting due to their capacity to combine fixed and random effects. This approach captures both broad trends and nuanced variations within age groups, genders, and countries. By accounting for data correlations and temporal dependencies, LME models offer an improved and more accurate analysis of mortality trends, leading to enhanced predictions and valuable insights for understanding and forecasting mortality patterns.

The LME model utilizes log ASDRs plotted against the mortality covariate $k_t$ to explore mortality variations, particularly across country-gender aggregates and within 24 distinct age groups. While the LME model offers valuable insights, it has its limitations. Notably, the relationship between $y_{cgxt}$ and $k_t$ in the LME model did not yield a perfect fit. To address this limitation, the authors introduces a GAMM (see \cite{wood2017generalized}). The GAMM aims to combine the strengths of the LME model while addressing its shortcomings by effectively capturing any nonlinear relationships between log ASDRs and the mortality covariate $k_t$. \cite{hilton2019projecting} and \cite{hall2011mortality} have previously employed generalized additive modeling in the projection of mortality rates. This approach has demonstrated its utility and effectiveness in similar contexts.

GAMMs can be considered a generalization of LME models, with the addition of the flexibility to model nonlinear relationships using additive functions (see \cite{wood2017generalized}). LMMs are a type of statistical model used to analyze data with both fixed effects (variables we are interested in studying) and random effects (variables that are not of primary interest but are included due to their impact on the data). LME models assume a linear relationship between the response variable and predictor variables. They are commonly used in various fields, including social sciences, biology, and economics. The fixed effects are modeled linearly, and the random effects account for variability due to unobserved factors. GAMMs extend the concept of LME models by allowing for the inclusion of nonlinear relationships between predictors and the response variable. They can handle not only linear but also nonlinear effects of predictors on the response. This is achieved through the use of additive functions, which means that each predictor can have a separate, flexible smooth function that captures its effect on the response. These smooth functions are often implemented using techniques like splines. GAMMs encompass the capabilities of LMMs by incorporating random effects and adding the flexibility to model nonlinear relationships using smooth functions. Therefore, we can view GAMMs as a generalization of LME models that provides a broader scope of modeling options. This added flexibility makes GAMMs especially useful when relationships between variables are not strictly linear and may exhibit more complex patterns.

The examination of ASDR trends concerning the mortality covariate $k_t$ unveils evident non-linear patterns. To effectively capture these intricacies, we employ GAM as our approach for analyzing mortality data. GAM, initially introduced by Hastie and Tibshirani (\citealt{hastie1990generalized}) and further refined by Wood in 2006, is a versatile regression technique that accommodates both linear and non-linear relationships between the dependent variable (ASDRs) and the independent variables (mortality covariate $k_t$ and $k_{ct}$). In contrast to traditional linear mixed effects methods, GAMMs provide us with the capability to model intricate interactions and dependencies between ASDRs and $k_t$, as well as between ASDRs and $k_{ct}$, resulting in more accurate modeling. This enhanced flexibility allows for a comprehensive understanding of the intricate associations within our data, leading to more precise and insightful conclusions in our study.

\section{Modeling Mortality Rates in Multiple Populations}\label{sec2}

\begin{figure}[H]
    \centering
    \includegraphics[width=\textwidth]{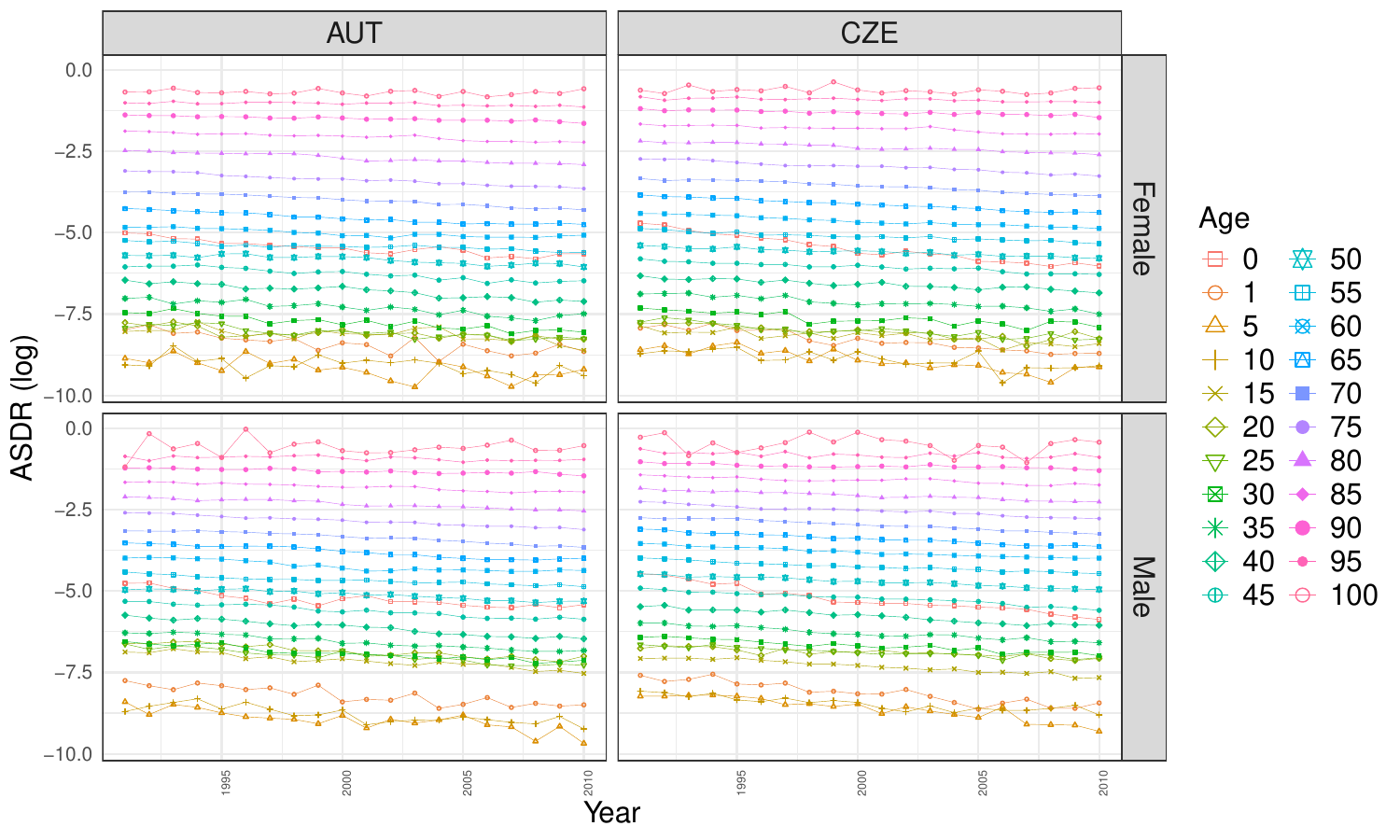}
    \caption{ASDRs across four populations from 1991 to 2010.}
    \label{fig4.1}
\end{figure}

Age-specific mortality patterns provide a unique window into the complex web of factors influencing mortality rates among various age groups. Think of these patterns as a kind of "mortality fingerprint," offering insights into the specific health challenges and vulnerabilities faced by people of different ages. While numerical analyses provide valuable insights, the art of data visualization unveils an additional layer of complexity and clarity. To fully grasp how mortality is distributed across different age groups within a population, it is crucial to examine mortality data carefully. Let us consider Figure \ref{fig4.1} as an example, where ASDRs are portrayed over a span of five decades for two countries. Each curve, shown on a logarithmic scale, tells a unique story of mortality over time. What becomes evident is a compelling tale of progress, as we witness a consistent decline in mortality rates, suggesting that, on average, people are enjoying longer lives. What is most fascinating, though, is the remarkable similarity in how mortality rates change, both within specific age groups and across diverse populations. Even when comparing different regions and demographics, a surprising consistency in the trajectory of mortality rates is found. Yet, amidst this regularity, the red curves at the center of Figure \ref{fig4.1}, representing infant mortality, stand out clearly, displaying significant fluctuations. This finding highlights the profound impact of infant health on the broader landscape of mortality. In essence, these age-specific patterns and temporal trends go beyond just mortality statistics; they tell a story of human resilience and adaptation in the face of evolving health challenges. By finding the regularity in both age patterns and trends over time, we conclude that the population age structure reveals a consistent and robust internal pattern, indicative of the broader health and societal factors at play in shaping longevity and mortality rates. Understanding these patterns extends beyond mere data analysis; it is an exploration of the human experience and the strategies societies employ to ensure longer, healthier lives.

\begin{figure}[H]
    \centering
    \includegraphics[width=\textwidth]{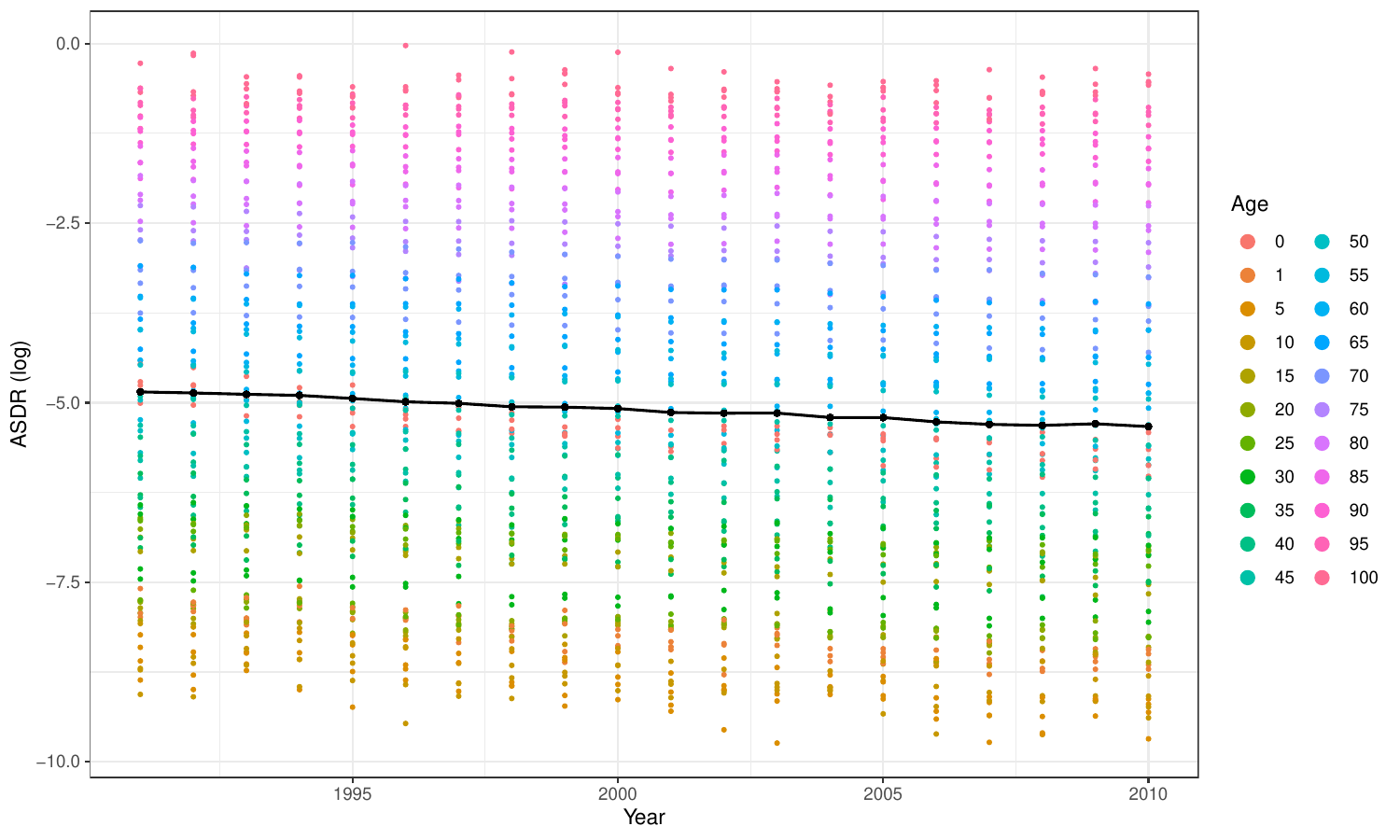}
    \caption{Average log mortality rates ($k_t$) across the four populations and all age groups (1991-2010). The thick black curve in this plot represents $k_t$, which is the average of log mortality rates across the four distinct populations and all age groups spanning the period from 1991 to 2010. $k_t$ serves as a comprehensive measure of the overall mortality trend during this two-decade interval, offering valuable insights into the collective impact on mortality rates across diverse populations and age groups.}
    \label{fig4.2}
\end{figure}

In Figures \ref{fig4.2}, a distinct insight emerges through the representation of the thick black curve, symbolizing $k_t$. This curve is derived from the logarithmic average of mortality rates spanning various age groups. Specifically, for instance, $k_{1961}$ reflects the average ASDRs for the year 1961. This aggregated curve, $k_t$, essentially captures the overarching trajectory of age-specific mortality rates within the dataset encompassing four different populations. However, a closer examination of the data reveals a noteworthy observation: the trends in mortality rates exhibit substantial diversity across age groups within individual countries. This disparity becomes evident when comparing the ASDRs against the average curve (the black curve), as they demonstrate significant deviations from each other.

Interestingly, the disparities within individual countries stand in contrast to the striking similarity observed in mortality rates among the same age groups across different populations. This pattern underscores the presence of consistent trends, irrespective of national boundaries. The commonality in mortality rate patterns at corresponding ages highlights a shared underlying factor, possibly linked to universal factors like biological aging processes or common health influences. This intriguing finding points toward the influence of intrinsic, age-related factors on mortality rates, which hold sway across various populations.

The method of GAMMs is now well established. Let $m_{c,g,x,t}$ denote the ASDR at age $x$ and time $t$ of gender $g$ in country $c$, for $c=$ $1, 2,\cdots,M$; $g=1, 2$; $x = 0, 1, \cdots , \omega$; and $t=0,1, \cdots, T$. Let $y_{cgxt}= \log (m_{c,g,x,t})$. Let $k_{ct}$ represent the average of $y_{cgxt}$ for two distinct age range groups: the average for age groups 0 to 40 and the average for age groups 41 to $\omega$, encompassing both females and males in country $c$ at time $t$. Additionally, let $k_t$ denote the average of $y_{cgxt}$ at time $t$ across all age groups in all countries and genders:
\begin{align}
k_{t} = \dfrac{\sum\limits_{c=1}^{M}\sum\limits_{g=1}^{2}\sum\limits_{x=0}^{\omega} y_{cgxt}}{2 M(\omega+1)},
\label{ali6}
\end{align}
for $c=1, 2, \cdots, M$, and $t=0, 1,  \cdots, T$.

\begin{figure}[H]
    \centering
    \includegraphics[width=\textwidth]{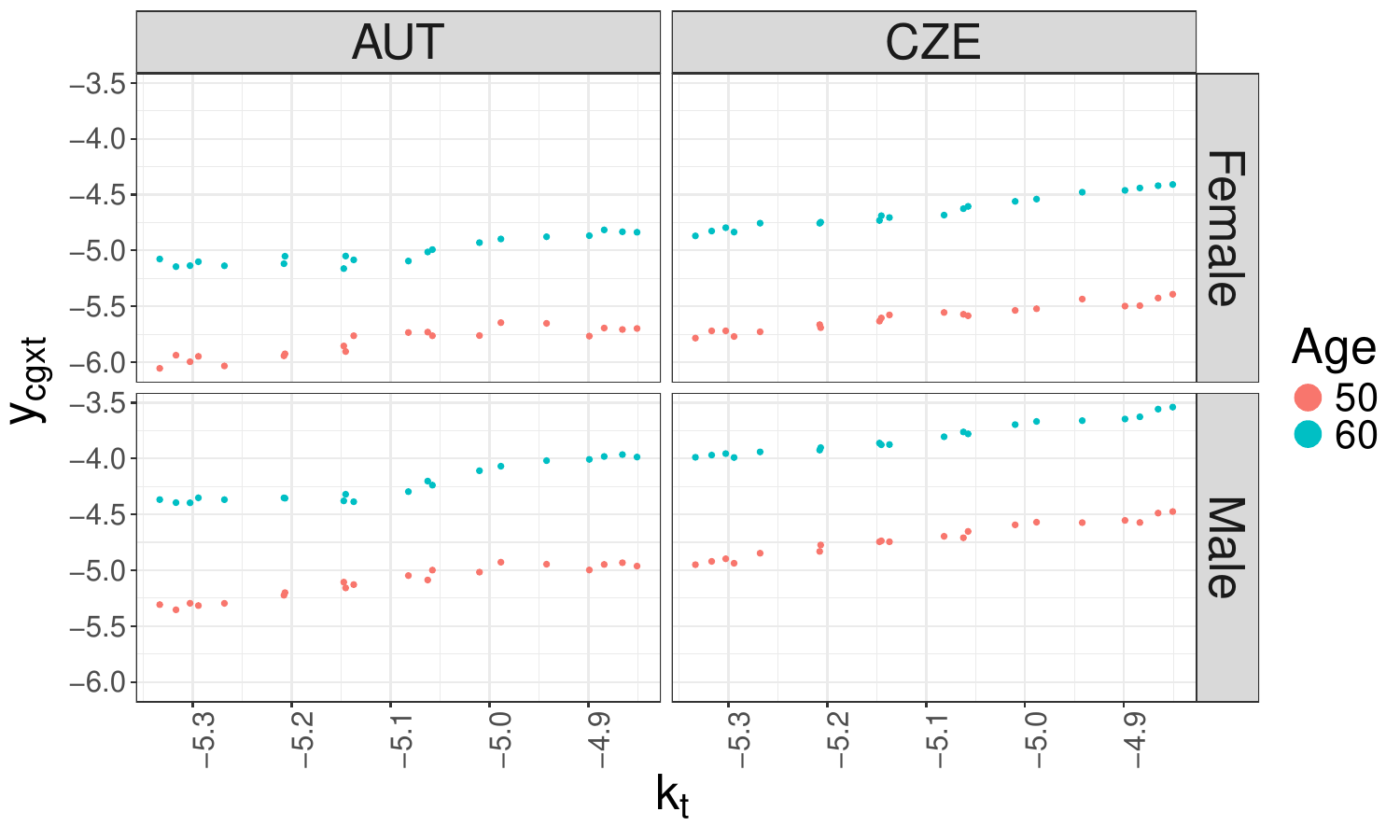}
    \caption{Exploring the relationship between logarithmically transformed ASDR trajectories ($y_{cgxt}$) and the mortality covariate ($k_t$) across the four populations (1991-2010). This plot investigates the connection between $y_{cgxt}$, representing logarithmically transformed ASDR trajectories, and the mortality covariate $k_t$ across the four populations during the period of 1991-2010. These trajectories reveal significant nonlinearity, underscoring the complexity of the association between log mortality and the covariate $k_t$.}
    \label{fig4.3}
\end{figure}

Figure \ref{fig4.3} displays logarithmically transformed ASDR trajectories, labeled as $y_{cgxt}$, in relation to the mortality covariate $k_t$. These trajectories are observed across distinct age groups, specifically  50 and 60, within each of the four populations that constitute the focus of our study. The plot illustrating the relationship between the response variable and the covariate is central to our approach in this chapter. Our capacity to achieve superior fitting within this plot directly results in reduced residuals, thereby enhancing the modeling and forecasting ASDRs. Furthermore, when introducing alternative statistical models for the ASDRs dataset, it is paramount to consider it as an initial step to assess how well the proposed model can capture the relationship between $y$ and $k_t$.

In the analysis of mortality data, it has become evident that the logarithmically transformed ASDR trajectories exhibit pronounced nonlinearity concerning the mortality covariate $k_t$ (see Figure \ref{fig4.3}). This inherent nonlinearity in the relationship between the dependent variable, log ASDR, and the independent variable, the mortality covariate, presents a complex analytical challenge. Traditional linear models struggle to accommodate these complexities, necessitating the adoption of a more sophisticated approach. To address this challenge, we have chosen to employ GAMMs. GAMMs are a powerful extension of the conventional GAMs framework, enriched by the incorporation of random effects. This augmentation equips GAMMs with the unique capability to effectively handle scenarios involving correlated and clustered responses. GAMMs excel in capturing nuanced relationships within the data, particularly when dealing with structured dependencies found in longitudinal data and research designs that incorporate repeated measurements. Their adaptability sets them apart, allowing us to move beyond the limitations of linear regression. Unlike traditional linear regression models, which are confined to modeling linear associations between the dependent variable ($y_{cgxt}$) and the independent variable ($k_t$), GAMMs proficiently accommodate non-linear associations. This adaptability is pivotal as it enables us to encapsulate intricate correlations and interactions intrinsic to the dataset.

\subsection{Model Formulation}

The command to fit a GAM for an individual age group using the bam (\cite{wood2015generalized}) function from the mgcv package can be stated as follows:

\begin{lstlisting}[caption={A GAMM for ASDRs in multiple populations.}, label=list:ex]
model = bam(y ~ age+gender:age+
                s(kt,bs="ts")+
                s(country:gender:age,bs="re") + 
                s(kt,country:gender:age,bs="fs",m=1),
                data=ASDRs)
\end{lstlisting}
The interpretation of this specific GAMM specification is as follows:

The age term in the model serves as an intercept for different age groups. It captures the baseline or constant effect of each age category on the response variable when all other factors are held constant. It provides a reference point for interpreting the impact of age categories. 

The gender:age term plays a vital role in capturing intercepts for each age group and, within each age group, for each gender category. This term allows for distinct baseline values for the response variable based on both age and gender. It identifies and accounts for constant (intercept) differences among these age groups and gender combinations. 

s($k_{t}$, bs="ts"): this term uses a thin-plate spline (bs="ts") to model the relationship between the predictor variable $k_{t}$ and the response variable $y_{cgxt}$. The thin-plate spline is a non-linear and low-rank smoother that allows for flexible and adaptive modeling of the relationship. It captures complex and non-linear patterns in the data without requiring the specification of knots. The smoothing parameter controls the degree of smoothness, allowing the model to adapt to the data patterns. 

s(country:gender:age, bs="re"): this term models a random effect for the combination of factor variables country, gender, and age using the "re" (random effects) basis. It captures random variability in the relationship between the response variable $y_{cgxt}$ and these factor combinations. This term allows for variation in intercepts among different levels of the factor combinations. 

s($k_{t}$, country:gender:age, bs="fs", m=1): this term models a factor-smooth interaction for the variable $k_{t}$ while considering the combination of country, gender, and age. It allows the smooth effect of $k_{t}$ to vary across different levels of this combination while maintaining a common smoothness constraint. 

In GAMMs, smooth functions are a crucial component that allows for the modeling of nonlinear relationships between predictors and the response variable. These smooth functions are used to capture complex and flexible patterns in the data that may not be well-modelled using simple linear relationships. Smooth functions are implemented using various mathematical techniques, and one common approach is through the use of splines. While splines are a common technique to implement smooth functions, there are other approaches as well:
B-splines are piecewise polynomial functions that are defined over intervals (segments). The overall smooth function is constructed by combining multiple B-splines with different coefficients at each segment. The coefficients control the shape of the curve, and they are estimated from the data. Natural splines are Similar to B-splines, but they impose additional constraints at the endpoints to ensure smoothness. This can be particularly useful when we want a smooth curve that does not oscillate too much at the edges. P-splines (penalized splines) combine the flexibility of splines with a regularization term that helps prevent overfitting. They use a penalty term on the coefficients of the spline to control the smoothness of the curve. This approach allows for automatic selection of the number and placement of knots while avoiding excessive wiggles in the curve. Smoothing splines are similar to B-splines but are obtained by minimizing a penalized residual sum of squares. They find the smoothest curve that fits the data within certain constraints. Thin plate splines (\cite{wood2003thin}) are a type of spline function that can be used to create smooth surfaces in multidimensional space. They are particularly useful when we have multiple predictor variables and want to capture complex interactions and nonlinear relationships among them.
Gaussian processes are a probabilistic framework that models the relationship between variables by assuming that any set of variables follows a joint Gaussian distribution. They allow for very flexible modeling of nonlinear relationships and can capture uncertainties in predictions. Local regression methods, like LOESS (Locally Weighted Scatterplot Smoothing), fit a separate regression function to each data point by giving more weight to nearby points. This allows for capturing local trends in the data. These techniques provide various ways to implement smooth functions in GAMMs. The choice of technique depends on factors like the complexity of the data, the desired smoothness of the curves, computational efficiency, and the specific goals of the analysis. The overarching goal is to capture the underlying patterns in the data while avoiding overfitting and achieving good generalization to new data. For one-dimensional predictors, such as continuous variables, the default smooth function technique in mgcv often employs cubic regression splines. Cubic splines are piecewise-defined cubic polynomials that are joined together at specific points called knots. These knots are chosen automatically based on the data distribution. The cubic nature of the splines ensures smoothness up to the second derivative, resulting in smooth curves that can capture nonlinear relationships. For higher-dimensional predictors or interactions between predictors, mgcv uses tensor product smooths. Tensor product smooths allow for the creation of smooth surfaces in multidimensional predictor spaces. These surfaces capture interactions and nonlinear relationships among the predictors. The smoothing is done in a way that ensures smoothness across all dimensions, while avoiding overly complex and wiggly curves

When using the mgcv package (\citealt{wood2015package}) in R to fit GAMs or GAMMs, the default smooth function technique is based on cubic regression splines for one-dimensional predictors and tensor product smooths for higher-dimensional predictors. The package uses these techniques to create smooth functions that capture the relationships between predictors and the response variable.

\begin{figure}[H]
    \centering
    \includegraphics[width=\textwidth]{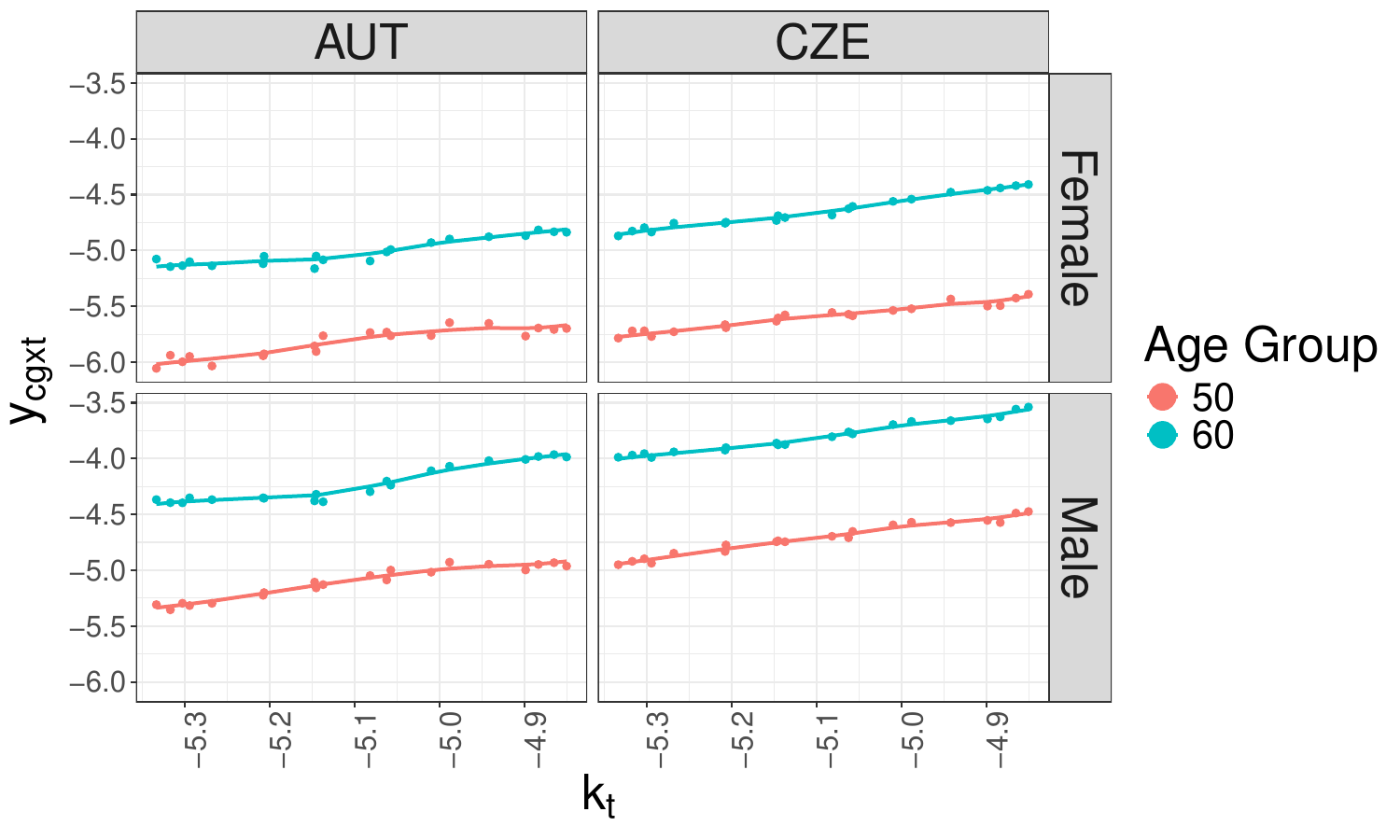}
    \caption{GAMM fits: $y_{cgxt}$ versus $k_{t}$. This plot highlights the GAMM's ability to effectively capture nonlinear patterns.}
    \label{fig4.4}
\end{figure}

Figure \ref{fig4.4} provides a compelling visual representation of the model's performance in capturing the complex relationship between log ASDRs and the mortality covariate $k_{t}$). The most striking aspect of this figure is the remarkable agreement between the fitted values and the observed values of log ASDRs. The closeness of the fitted values to the observed values illustrates the model's ability to capture the underlying patterns in the data accurately. The figure clearly exhibits nonlinearity in the relationship between log ASDRs and $k_{t}$. Unlike a simple linear model, which may have struggled to capture this nonlinearity, the GAMM approach excels in accommodating and modeling the intricate non-linear patterns. The curves and deviations in the plot showcase the richness and complexity of the relationship. The figure serves as a compelling justification for employing the GAMM methodology. The pronounced nonlinearity in the data is vividly captured by the model, highlighting the necessity of using a flexible approach like GAMM. A simpler linear model would have likely fallen short in capturing these intricate patterns. The figure further reveals that the relationship between log ASDRs and $k_{t}$ differs across gender and between the two countries in the study. The separate curves for each gender and country underline the importance of accounting for these interactions and variations in the model. The agreement between fitted and observed values is not just visual; it is also statistically supported. This figure reinforces the strong statistical significance of the GAMM terms related to $k_{t}$ and the associated non-linear relationships. Therefore, this figure demonstrates the power of the GAMM approach in modeling and capturing the nonlinearity in the relationship between log ASDRs and the mortality covariate $k_{t}$. The close alignment between fitted and observed values affirms the model's accuracy, while the presence of nonlinearity emphasizes the need for flexible modeling techniques like GAMM in complex data analysis.

Table~\ref{tab4.1} presents a comparison of MSE between two models, the GAMM and the LL model~\cite{li2005coherent}, for forecasting mortality across four populations. The results show that the GAMM outperforms the LL model in both in-sample and out-of-sample accuracy for all four populations. This highlights the GAMM’s effectiveness in producing more accurate mortality forecasts. Its superior performance suggests that the GAMM captures nuanced patterns and complexities in the data that the LL model may overlook, leading to improved predictive accuracy. These findings have practical implications in fields such as healthcare and policy-making, where reliable mortality forecasts are essential. The consistent advantage of the GAMM demonstrates the value of advanced modeling techniques in leveraging complex data structures.

\begin{table}[ht]
\centering
\caption{MSE for GAMM and LL models on testing sets across four populations}
\label{tab:newtest}
\begin{tabular}{lcccc}
\toprule
\textbf{Model} & \textbf{AUT Females} & \textbf{AUT Males} & \textbf{CZE Females} & \textbf{CZE Males} \\
\midrule
LL   & 0.037883 & 0.038207 & 0.028665 & 0.018618 \\
GAMM & 0.023031 & 0.017990 & 0.019246 & 0.017727 \\
\bottomrule
\end{tabular}
\label{tab4.1}
\end{table}

\bibliographystyle{chicago}
\bibliography{mybib}

\end{document}